\documentclass[english]{article}
\usepackage[T1]{fontenc}
\usepackage[utf8]{inputenc}
\usepackage{amsmath}
\usepackage{amssymb}
\usepackage{graphicx}
\usepackage{esint}
\usepackage{algorithm,algorithmic}

\usepackage{xcolor}

\makeatletter
\usepackage{babel}

\usepackage{setspace}
\@ifundefined{showcaptionsetup}{}{%
 \PassOptionsToPackage{caption=false}{subfig}}
\usepackage{subfig}
\makeatother

\begin{document}

\title{Multiclass histogram-based thresholding using kernel
density estimation and scale-space representations}

\author{S. Korneev, J. Gilles, I. Battiato}

\maketitle

\begin{abstract}
We present a method for multiclass thresholding of histograms whose modes may overlap so strongly that they merge into a unimodal distribution. The histogram is approximated by a nonparametric kernel density (KD) estimate, a mixture of Gaussians with one component per bin and a common bandwidth, whose parameters are determined by Expectation-Maximization (EM) iterations. Because the estimate is a Gaussian mixture, the semigroup property of Gaussian convolution yields its scale-space representation in closed form, at bandwidths both larger and smaller than the fitted one. The algorithm increases or decreases the bandwidth until the number of local minima of the estimate equals the number of requested classes minus one, and returns the positions of those minima as the threshold values. We validate the method on synthetic histograms with known thresholds, and we apply it to the histogram of X-ray computed tomography (XCT) images of a natural porous sample in which the three phases strongly overlap. The porosity computed from the resulting segmentation agrees with the direct experimental measurement to within $4\%$, whereas K-means clustering misestimates it by $52\%$.
\end{abstract}

\section{Introduction}

A histogram-based thresholding algorithm offers several advantages over direct clustering of the underlying data for classification tasks: it is simple to understand and it is
computationally efficient once the histogram has been obtained~\cite{Glasbey1993}. Histogram-based
methods can be classified into two main categories: parametric and nonparametric. Parametric methods, such as those based on the EM algorithm, estimate the parameters of a given model of the histogram distribution; parametric methods using mixtures of generalized Gaussian distributions were proposed in~\cite{Bazi2007, Boulmerka2014}. Nonparametric methods instead threshold the histogram according to a predetermined criterion~\cite{Glasbey1993}, and they
have proven to be robust and more accurate than parametric ones in a number of instances, such as high-dimensional data and clusters that deviate substantially from Gaussian distributions~\cite{ROBERTS1997261, Bruce2007}. Classical criterion-based methods include K-means clustering, Otsu's between-class variance criterion~\cite{Otsu1979} together with its multilevel extensions~\cite{Liao2001}, and the minimum error thresholding of Kittler \& Illingworth~\cite{Kittler1986}, which thresholds the histogram by minimizing the Bayes classification
error under the assumption of a bimodal Gaussian distribution. The minimum error approach was later generalized to mixtures of not-well-separated distributions~\cite{Cho1989}, and a recent Bayesian formulation unifies Otsu's method and minimum error thresholding within a single family~\cite{Barron2020}. A comparative analysis of various nonparametric methods can be found in~\cite{Nacereddine2005,Glasbey1993}.

As the number of thresholds increases, the optimization
problem solved in such methods can become computationally intensive
and unstable. Some of these issues can be overcome by histogram shape-based thresholding. In this regard, Rosin~\cite{Rosin2001} suggested defining the thresholds by finding ``corners'' in the histogram plot. Shape-based methods require the threshold values to be clearly reflected in the histogram's shape, yet noise and bin-width variations may affect their accuracy and stability. To overcome these issues, various filters can be applied to smooth the histogram. Tsai~\cite{Tsai1995} proposed to find the maximum curvature of the histogram plot after applying a Gaussian filter. In the method developed by Gilles \& Heal~\cite{Gilles2014}, the optimal thresholds are found by tracking local minima through a scale-space representation, based on the observation that meaningful threshold values correspond to local minima that persist through the scales. The connection between the kernel bandwidth and the number of modes of a density estimate is classical in the statistics literature: Silverman~\cite{Silverman1981} tested multimodality using the critical bandwidth, i.e., the smallest bandwidth at which a kernel density estimate has at most $m$ modes, and Minnotte \& Scott~\cite{Minnotte1993} visualized the positions of the modes across bandwidths as a ``mode tree''. Thresholding along persistent local minima inherits the advantages of
parametric methods while being computationally more stable and efficient:
it approximates the minimum-Bayes-error thresholds
without explicitly identifying the distribution parameters, since it
only requires that the clusters are separated by meaningful
local minima. However, this requirement may not always be met, specifically when the component
distributions are not well separated and merge into a unimodal histogram.

Highly overlapping classes are frequent in low-contrast image thresholding
problems~\cite{Kopriva2015}, for instance in images produced by
gradient-based edge detection algorithms~\cite{Medina-Carnicer2008},
in medical X-ray computed tomography (XCT)~\cite{Tavares2011}, in computer
vision~\cite{Huang2010,Chen2012}, and in security and surveillance at low illumination
and low exposure~\cite{LiangshengWang2009,Lore2017,Huang2008}.
Popular edge detection methods are based on thresholding gradient maps,
i.e., on segmenting the gradient-map histogram into two clusters; such maps have low contrast and unimodal
histograms with long tails~\cite{Medina-Carnicer2008}. The clusterability of highly overlapping clusters is a subject of discussion~\cite{Hennig2015}, and recent work argues that highly overlapping clusters that fail a multimodality test should not be clustered at all~\cite{ADOLFSSON201913}. In the applications considered here, however, the number of classes is known \textit{a priori} from the physics of the problem.

Nevertheless, histogram-based methods have been proposed to threshold unimodal histograms into two clusters. For example, the threshold can be obtained by piecewise linear regression, finding the two segments that best fit the descending slope of the
histogram~\cite{Coudray2010}. Another method assigns a 2D point in a
receiver operating characteristic (ROC) space to each possible threshold of the histogram, without a prior reference image, and determines the optimal threshold from the ROC graph~\cite{Medina-Carnicer2008}. Although these methods are efficient for binary segmentation of unimodal distributions, they are difficult to extend to the multiclass thresholding case. We refer the reader to the review papers~\cite{Baradez2004,Rosin2001} on nonparametric thresholding of unimodal distributions for more details. In this work, we propose a new multiclass thresholding method that thresholds along the local minima and is able to handle highly overlapping distributions, including histograms whose modes have merged into a unimodal shape.

In Section~\ref{sec:Problem-formulation}, we provide some generalities regarding histogram modeling and we review scale-space representations and their basic properties. In Section~\ref{sec:Thresholding-algorithm}, we introduce the three main steps of the thresholding algorithm.
In Section~\ref{sec:Thresholding-synthetic}, we validate the proposed algorithm using a set of synthetic histograms with known threshold values. In Section~\ref{sec:Thresholding-XCT}, we threshold low-contrast tomographic images of a natural porous sample. The obtained segmentation allows one to estimate the sample porosity, which we compare with direct experimental measurements. Finally, we summarize our work in Section~\ref{sec:Conclusion}.

\section{Generalities}\label{sec:Problem-formulation}
\subsection{Mixture models of histograms}
Let us define a histogram $\mathcal{H}$ as the set of values given by
\begin{equation}
\mathcal{H}=\left\{ \left\{ t_{i},\ h_{i}\right\} \right\} _{i=1}^{N}\quad\text{s.t.}\quad\Delta t\sum_{i=1}^{N}h_{i}=1,\label{eq:histogram}
\end{equation}
where $h_{i}$ corresponds to the normalized number of occurrences in the bin of width $\Delta t$
centered at $t_{i}$ (all bins have the same width). A histogram-based segmentation algorithm identifies thresholds that split $\mathcal{H}$ into $C$ clusters, where $C$ is a fixed input parameter. To formulate this thresholding problem, we assume that the histogram \eqref{eq:histogram} corresponds to a sampled version of a mixture of $K$ unimodal distributions (or modes), $H_k,\ k=1,\ldots,K$, with $K=C$, i.e.,
\begin{equation}
H(t)=\sum_{k=1}^{K}\alpha_{k}H_k(t) \quad\text{s.t.}\quad\sum_{k=1}^{K}\alpha_{k}=1,\label{eq:mixture-model}
\end{equation}
where $\alpha_{k}$ is the \textit{a priori} probability of occurrence of the $k$-th component and $t$ is a continuous variable. Several choices of unimodal component are possible; here, we use Gaussian functions, i.e.,
\begin{align}
 H_k(t)=\mathcal{G}\left(t;\mu_k,\sigma_k^{2} \right),
 \end{align}
where $\mathcal{G}(t;\mu,\sigma^2)=(2\pi\sigma^2)^{-1/2}\exp\left[-(t-\mu)^2/(2\sigma^2)\right]$ denotes the Gaussian density with mean $\mu$ and variance $\sigma^2$.

For any given pair of consecutive modes $\left\{H_{k},H_{k+1}\right\}$, we define the threshold value $\tau_{k}$ as the intersection point of the weighted modes that lies between their means, i.e., $\tau_k\in(\mu_k,\mu_{k+1})$ solves

\begin{equation}
\alpha_{k}H_{k}(\tau_{k})=\alpha_{k+1}H_{k+1}(\tau_{k}),
\label{eq:minimum-Bayes-error}
\end{equation}
when such a point exists; it is then unique, since the log-ratio of two Gaussian densities is strictly monotone between their means. In a neighborhood of $\tau_{k}$, the weighted mode $\alpha_{k}H_{k}$ dominates for $t<\tau_{k}$ and $\alpha_{k+1}H_{k+1}$ dominates for $t>\tau_{k}$, so $\tau_{k}$ is the optimal single threshold separating the two classes (with unequal variances, the full Bayes rule may place an additional boundary in the far tails).

In the following, we assume the modes $H_k$ are well-ordered, i.e., their mean values satisfy $\mu_{k+1}>\mu_{k}$, where the mean and the variance of $H_k$ are defined, respectively, by
\begin{align}
\mu_{k}=&\int_{-\infty}^{+\infty} t H_k(t) dt,\\
\sigma_k^2=&\int_{-\infty}^{+\infty} (t-\mu_{k})^2 H_k(t) dt.
\end{align}

\begin{figure}[t]
\begin{centering}
\subfloat[Components are resolved.\label{fig:schematic-mixture-model-resolved}]{\begin{centering}
\includegraphics[width=0.48\textwidth]{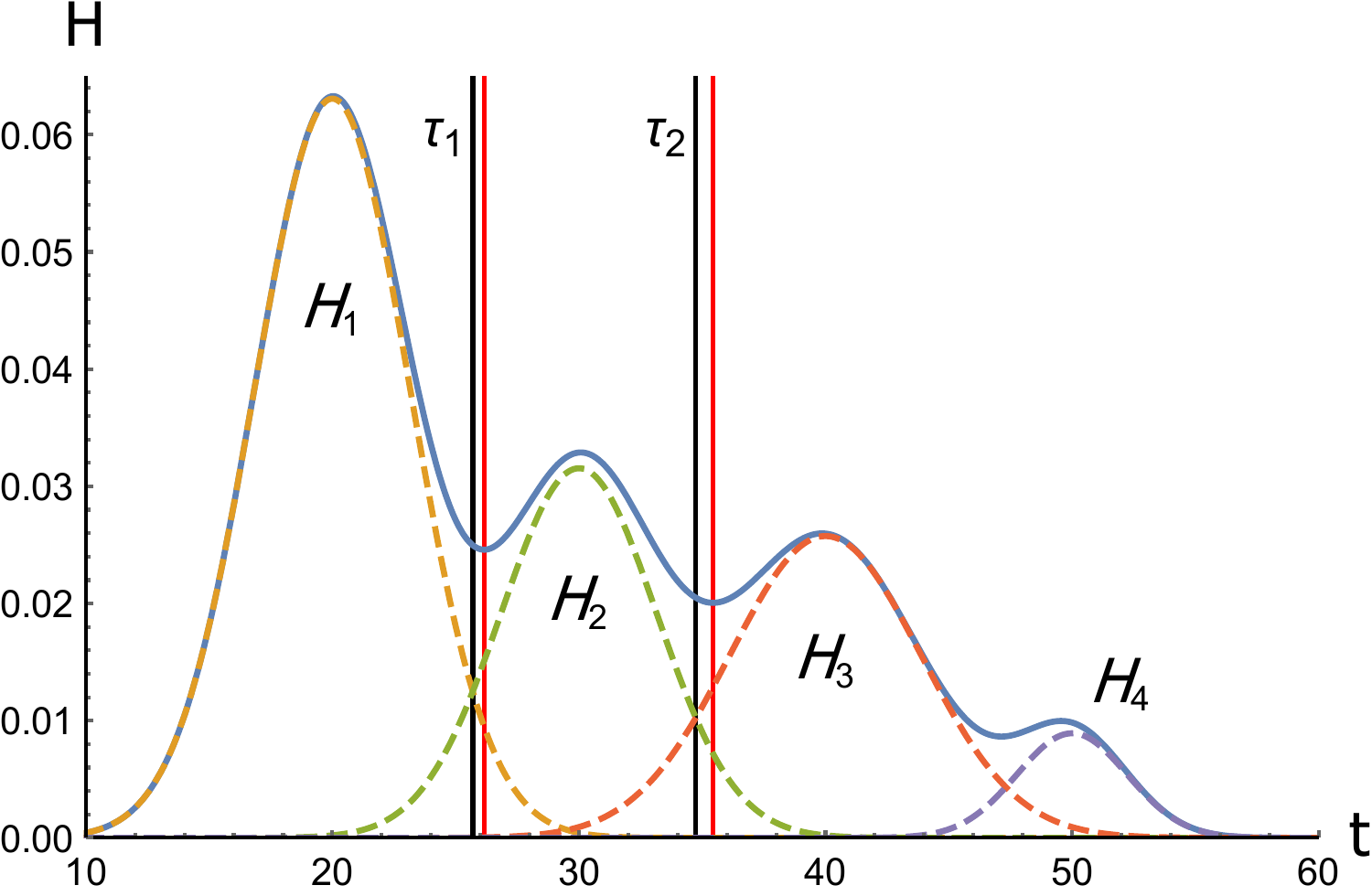}
\par\end{centering}
}\hfill{}\subfloat[Components are not resolved.\label{fig:schematic-mixture-model-notresolved}]{\begin{centering}
\includegraphics[width=0.48\textwidth]{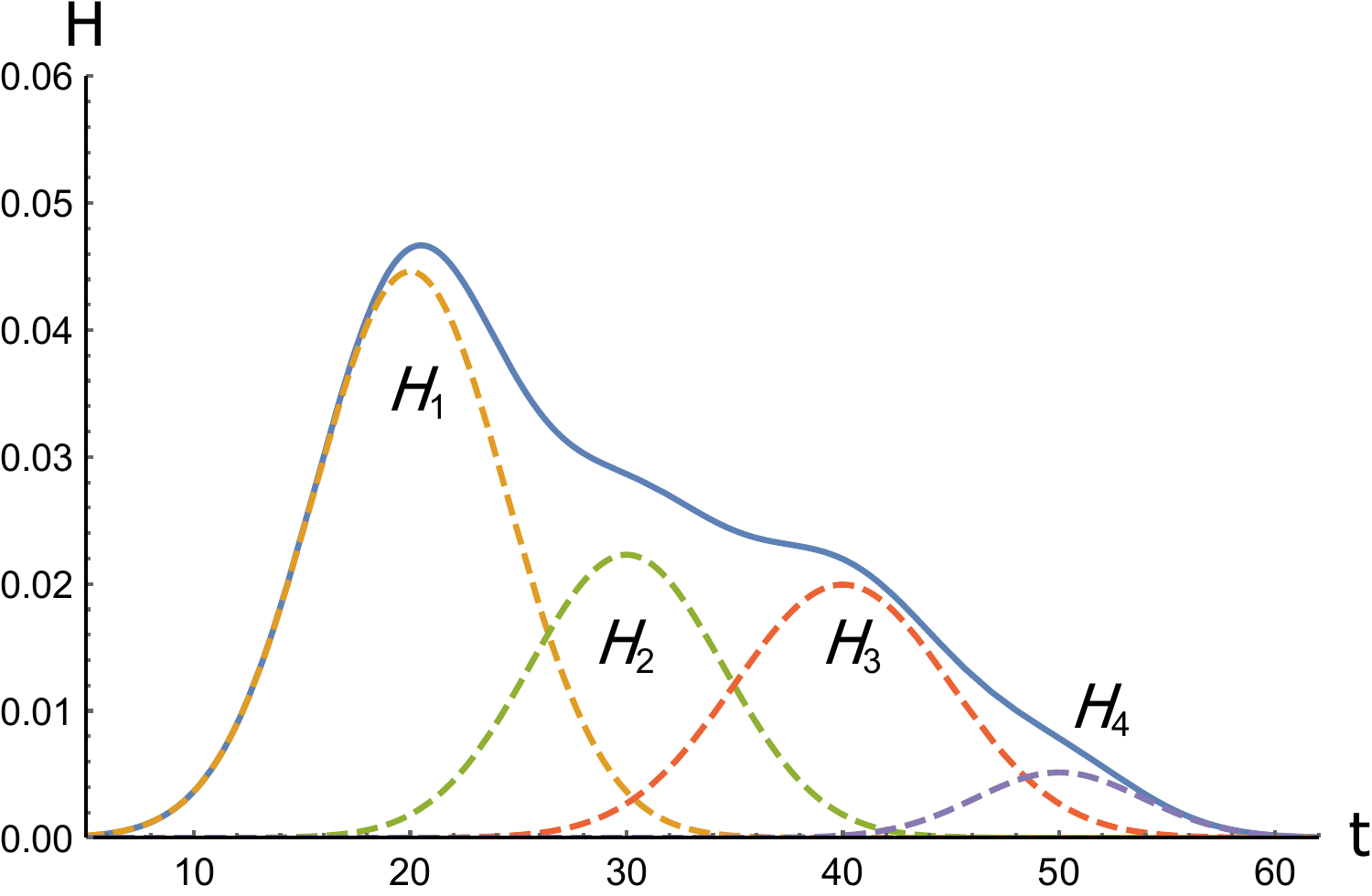}
\par\end{centering}
}
\par\end{centering}
\caption{Schematic representation of a mixture model with (a) well-separated
components and (b) highly overlapping components. The blue solid line shows the mixture model while
dashed lines correspond to components of the mixture.
The black vertical line shows the position of the component intersection
points and the red vertical line shows the closest local minimum of
the mixture.}
\end{figure}

If the modes are well separated, the threshold value $\tau_{k}$ is well approximated by the position of the local minimum of the mixture between the two modes (see Figure~\ref{fig:schematic-mixture-model-resolved}); for two modes with equal weights and equal variances, the local minimum and the intersection point coincide exactly, while in general the local minimum provides an approximation of $\tau_{k}$. In contrast, for highly overlapping modes, the intersection points cannot be resolved by finding local minima at all (see Figure~\ref{fig:schematic-mixture-model-notresolved}) and a different strategy must be adopted.

In Section~\ref{sec:Thresholding-algorithm}, we present the algorithm, which includes the reconstruction of a mixture of Gaussians, the detection of local minima and the identification of the appropriate thresholds.

\subsection{Scale-space representation}
Since our algorithm is based on the notion of scale-space representations, in this section we review how such representations are obtained as well as their basic properties. Given a
function $f(t)$, the scale-space representation of $f$ at scale $s$ is given by
\begin{align}
\mathcal{S}_f(t,s)=(f\ast\mathcal{G}(\cdot;0,s))(t),
\end{align}
where $\ast$ denotes convolution; here the scale $s$ is the variance of the Gaussian kernel. As the scale $s$ increases, this process progressively attenuates the high-frequency content of $f$. Moreover, in one dimension Gaussian smoothing possesses a causality property: no new local extrema are created as the scale increases~\cite{Babaud1986}, so the number of local extrema of $\mathcal{S}_f(\cdot,s)$ is a non-increasing function of $s$ and, for a large enough $s$, only a single maximum survives; for mixtures of Gaussians this property is analyzed in~\cite{Carreira2003}. This property is at the core of the method of Gilles \& Heal~\cite{Gilles2014}, which detects consistent local minima by tracking their ``long-life'' presence through the scales.

\section{Thresholding algorithm}\label{sec:Thresholding-algorithm}

The algorithm is composed of three steps. In Section~\ref{sec:KDE}, we describe the kernel density (KD) estimation of the histogram using Expectation-Maximization (EM) deconvolution with a Gaussian kernel. In Section~\ref{sec:minima}, we describe the extraction of the local minima of the KD estimate at a given scale. In Section~\ref{sec:method}, we describe the scale-space search that adjusts the scale until the required number of minima is obtained, which yields the thresholds. Representing the histogram by a mixture of Gaussians makes the scale search computationally efficient, thanks to the semi-group property of Gaussian functions, $\mathcal{G}(t;0,s_1)\ast\mathcal{G}(t;0,s_2)=\mathcal{G}(t;0,s_1+s_2)$.

\subsection{Kernel density estimation}\label{sec:KDE}
First, we approximate the histogram \eqref{eq:histogram} by a kernel density estimate consisting of a mixture of Gaussians,
\begin{equation}\label{eq:KD-estimation}
\text{KD}\left(t, \mathcal{B}, \sigma^2\right)=\sum_{j=1}^{N}\beta_{j}\mathcal{G}\left(t;t_{j},\sigma^{2}\right)\Delta t,
\end{equation}
where the nonnegative weights $\mathcal{B}=\left\{ \beta_{j}\right\} _{j=1}^{N}$ and the common variance $\sigma^{2}$ are unknown parameters to be determined. The variance $\sigma^{2}$ (the squared bandwidth of the kernel) is referred to as the scale in the scale-space nomenclature.

It is worth emphasizing that we associate one Gaussian with each bin, instead of assuming a small number of Gaussians whose count is generally not known in advance. This choice avoids the need to estimate the number of components in the mixture. Furthermore, since we assume that all bins have the same width, we use the same variance for each Gaussian in the mixture. Sharing a single variance has the additional advantage of avoiding the collapse of individual components onto single bins in the EM algorithm described hereafter.

We determine the parameters $(\mathcal{B},\sigma^2)$ by minimizing the Csisz\'ar I-divergence between the histogram and the model,
\begin{equation}
\underset{\mathcal{B}\geq 0,\ \sigma^2>0}{\min} \left \{ \Delta t\sum_{i=1}^{N} \left[h_i \ln\frac{h_i}{\text{KD}\left(t_i, \mathcal{B}, \sigma^2 \right)} - h_i + \text{KD}\left(t_i, \mathcal{B}, \sigma^2 \right) \right] \right\},\label{eq:minimization}
\end{equation}
with the convention $0\ln 0=0$. Minimizing \eqref{eq:minimization} corresponds to maximum-likelihood estimation when the bin counts follow a Poisson (or multinomial) distribution~\cite{Idier2008}. The joint estimation of $(\mathcal{B},\sigma^2)$ from \eqref{eq:minimization} is an ill-posed deconvolution problem~\cite{Idier2008,Burger2008}: as $\sigma^2\rightarrow 0$ the model can interpolate the histogram exactly (with $\beta_j\propto h_j$), so the objective alone does not determine the bandwidth. In practice, the problem is regularized implicitly by solving \eqref{eq:minimization} with early-stopped EM iterations: the iterations are initialized at the bin scale, $\sigma^{2(0)}=\Delta t^2$, and terminated when the change in the weights falls below a tolerance $\delta$, in the spirit of early-stopped Richardson--Lucy deconvolution~\cite{Burger2008}. The early stopping also regularizes against noise in the histogram.

Denoting the fast Fourier transform and its inverse by $\mathcal{F}$ and $\mathcal{F}^{-1}$, and writing
\begin{equation}
X\circledast Y=\Delta t\,\mathcal{F}^{-1}\left[\mathcal{F}\left[X\right]\mathcal{F}\left[Y\right]\right]
\end{equation}
for the linear discrete convolution of two vectors on the bin grid (the transforms are zero-padded to avoid circular wrap-around), we give the EM-KD iterations in Algorithm~\ref{eq:EM-iterations}. Here $\bar{\mathcal{H}}=\{h_i\}_{i=1}^N$ denotes the vector of bin heights, the kernel vectors $G^{(n)}$ and $V^{(n)}$ are sampled at the lags $l\Delta t$, $l=-(N-1),\ldots,N-1$, and all divisions and multiplications between vectors are pointwise.
\begin{algorithm}[!t]
\caption{Expectation-Maximization Kernel Density Estimation algorithm}
\label{eq:EM-iterations}
\begin{algorithmic}
\REQUIRE $\Delta t,\delta, \bar{\mathcal{H}}=\left\{ h_{i} \right\} _{i=1}^{N}$
\STATE Initialize: $\mathcal{B}^{(0)}=\left\{ 1/\left(N\Delta t\right)\right\} _{j=1}^{N},\ \sigma^{2(0)}=\Delta t^2$
\REPEAT
\STATE \begin{enumerate}
\item $G^{\left(n\right)}=\left\{ \mathcal{G}\left(l\Delta t;0,\sigma^{2\left(n\right)}\right)\right\} _{l=-(N-1)}^{N-1}$
\item $V^{\left(n\right)}=\left\{ \left(l\Delta t\right)^{2}\mathcal{G}\left(l\Delta t;0,\sigma^{2\left(n\right)}\right)\right\} _{l=-(N-1)}^{N-1}$
\item $D^{\left(n\right)}=\bar{\mathcal{H}}/\left(\mathcal{B}^{\left(n\right)}\circledast G^{\left(n\right)}\right)$
\item $\mathcal{B}^{\left(n+1\right)}=\mathcal{B}^{\left(n\right)}\left(D^{\left(n\right)}\circledast G^{\left(n\right)}\right)$
\item $\sigma^{2\left(n+1\right)}=\Delta
t\sum\mathcal{B}^{\left(n\right)}\left(D^{\left(n\right)}\circledast V^{\left(n\right)}\right)$
\end{enumerate}
\UNTIL{$\Delta t\sum\left|\mathcal{B}^{\left(n+1\right)}-\mathcal{B}^{\left(n\right)}\right|<\delta$}
\ENSURE $\mathcal{B},\sigma^2$
\end{algorithmic}
\end{algorithm}
For a fixed $\sigma^2$, steps 3--4 are exactly the Richardson--Lucy deconvolution iterations
\cite{Richardson1972a,Lucy1974}, and they conserve the total mass, $\Delta t\sum_j\beta_j^{(n)}=1$ for all $n$. Step 5 additionally re-estimates the common kernel bandwidth (it is the EM update of a variance shared by all components), so the complete scheme is an EM-type algorithm for a mixture with fixed centers, free weights, and a single shared variance. At a perfect fit, $D^{(n)}=1$ and step 5 leaves $\sigma^2$ unchanged (up to the discretization of the kernel).

\begin{figure}[!t]
\centering{}\subfloat[Initialization]{\centering{}\includegraphics[width=2.85in]{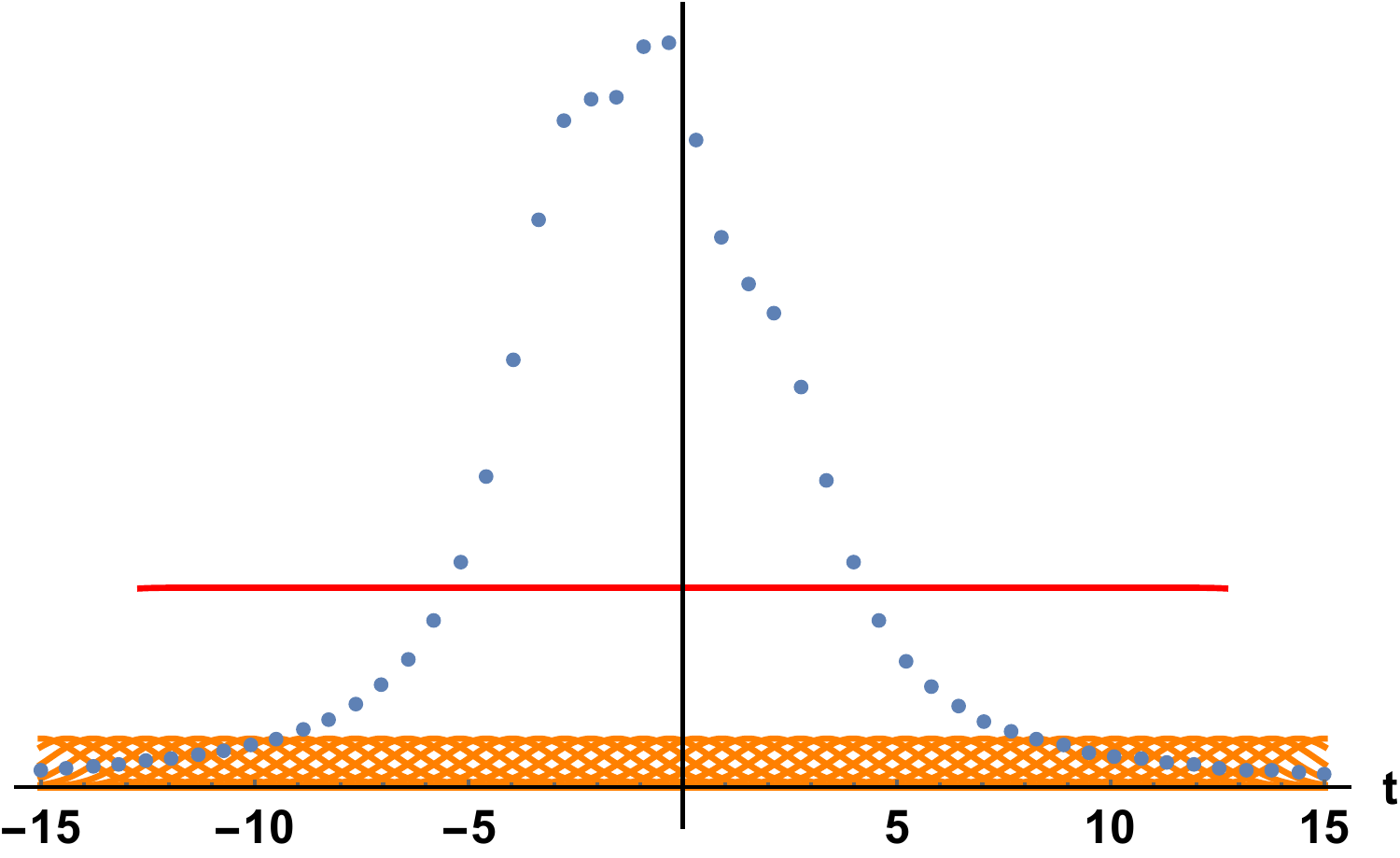}}
\hfill{}
\subfloat[Iteration 10]{\centering{}\includegraphics[width=2.85in]{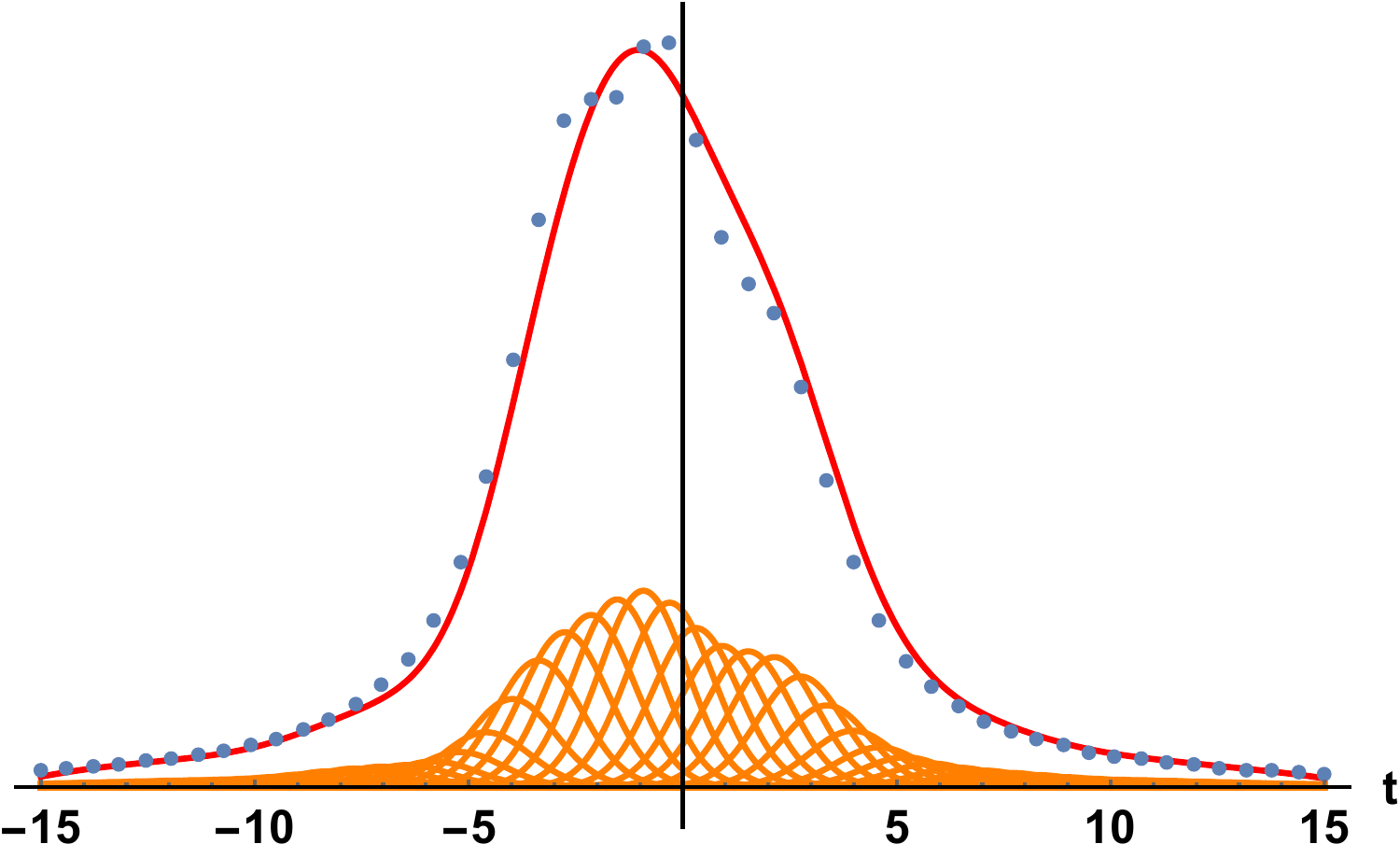}}
\hfill{}
\subfloat[Last iteration]{\centering{}\includegraphics[width=2.85in]{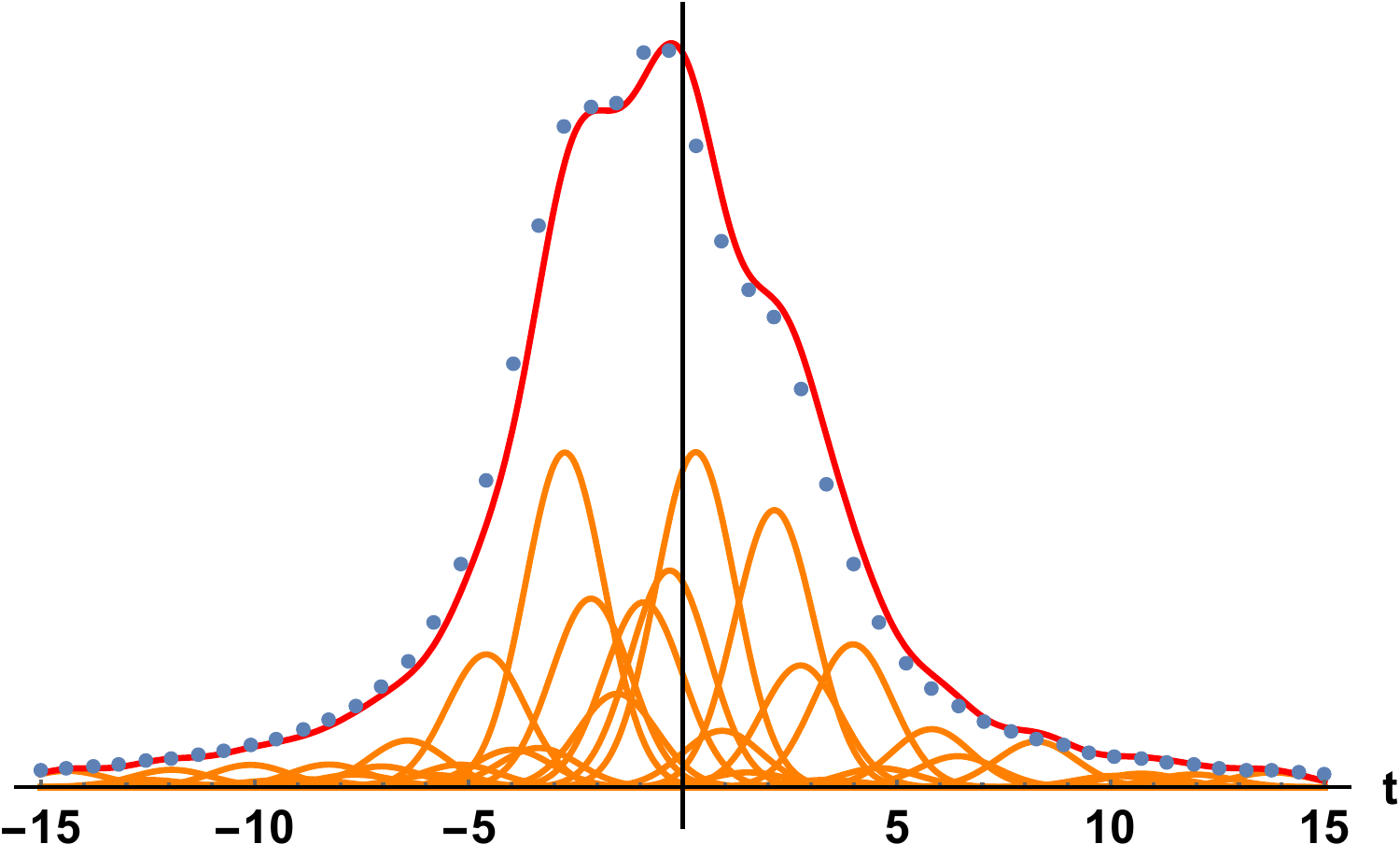}}
\caption{Evolution of the KD estimate of the histogram across the iterations of Algorithm~\ref{eq:EM-iterations}, where the blue dots show the histogram. The orange curves represent the
separated modes while the red curve
corresponds to the sum of the modes.}
\label{fig:steps}
\end{figure}
In Figure~\ref{fig:steps}, we show several iterations of the proposed KD estimation algorithm.

\subsection{Local minima detection}\label{sec:minima}
Given a mixture $\text{KD}(t,\mathcal{B},\sigma^2)$, we now need to extract the positions of all local minima. These minima can be detected in a semi-analytical fashion. Since an analytic form of $\text{KD}(t,\mathcal{B},\sigma^2)$ is available, its derivatives can be calculated and their zeros numerically determined. This procedure corresponds to numerically solving the following set of constraints for $t$:
\begin{subequations}
\begin{eqnarray}
\frac{d\text{KD}\left(t,\mathcal{B},\sigma^{2}\right)}{dt}=\sum_{j=1}^{N}\beta_{j}\frac{d\mathcal{G}\left(t;t_{j},\sigma^{2}\right)}{dt}\Delta
t=0\quad\text{s.t.}\\
\frac{d^{2}\text{KD}\left(t,\mathcal{B},\sigma^{2}\right)}{dt^{2}}=\sum_{j=1}^{N}\beta_{j}\frac{d^{2}\mathcal{G}\left(t;t_{j},\sigma^{2}\right)}{dt^{2}}\Delta
t>0
\label{eq:find-minima},
\end{eqnarray}
\end{subequations}
where the analytic forms of the derivatives are given by
\begin{equation}
\frac{d\mathcal{G}\left(t;t_{j},\sigma^{2}\right)}{dt}=-\mathcal{G}\left(t;t_{j},\sigma^{2}\right)\left( \frac{t-t_j}{\sigma^{2}}\right),
\end{equation}
and
\begin{equation}
\frac{d^2\mathcal{G}\left(t;t_{j},\sigma^{2}\right)}{dt^2}= \mathcal{G}\left(t;t_{j},\sigma^{2}\right)\left[ \frac{(t-t_j)^2-\sigma^{2}}{\sigma^{4}}\right].
\end{equation}
In the remainder of the paper, the set of local minima will be denoted by $\mathcal{M}$ and its cardinality by $\#\mathcal{M}$.

\subsection{Scale-space search for the thresholds}\label{sec:method}
In order to detect the expected thresholds, we use a scale-space representation of our histogram. We start with the scale-space representation of the histogram model, outlined in Section~\ref{sec:scale-space}, and then we present the overall threshold detection algorithm in Section~\ref{sec:algorithm}.

\subsubsection{Scale-space representation of the histogram model}\label{sec:scale-space}
Instead of building a scale-space representation of the original histogram $\mathcal{H}$, we take advantage of the analytical form provided by the KD estimation model. Specifically, since we use a mixture of Gaussians and scale-space representations in the continuous case are obtained by convolution with a Gaussian $\mathcal{G}(t;0,s)$, one can use the semi-group property of Gaussian functions: the scale-space representation $\mathcal{S}_{\text{KD}}(t,s)$ of a histogram $\mathcal{H}$, given its KD estimate $\text{KD}(t,\mathcal{B},\sigma^2)$, is simply
 \begin{align}
\mathcal{S}_{\text{KD}}(t,s)=\text{KD}(t,\mathcal{B},\sigma^2+s),\qquad s>-\sigma^2.
\end{align}
The advantage of having a continuous model is twofold. On the one hand, choosing $s>0$ yields the ordinary (smoothing) scale-space representation of the fitted model. On the other hand, choosing $s<0$ yields an ``inverse'' scale-space representation: the fitted mixture is evaluated at a bandwidth below the fitted scale, which corresponds to a partial Gaussian deconvolution of the fitted density. This inverse representation is meaningful inasmuch as the fitted mixture represents the histogram, and it is the key ingredient for dealing with overlapping distributions in the threshold detection algorithm presented in Section~\ref{sec:algorithm}. In the search below, the scale offset $s$ is varied in steps of magnitude $\Delta\sigma^2$, which sets the resolution of the scale search: smaller steps require more iterations but reduce the risk of stepping over the scale at which the required number of minima first appears.

\subsubsection{Threshold detection algorithm}\label{sec:algorithm}
Once the scale-space representation $\mathcal{S}_{\text{KD}}(t,s)$ is constructed, the expected thresholds that segment the histogram into $C$ clusters need to be identified. First, the set of local minima
$\mathcal{M}$ of $\text{KD}(t,\mathcal{B},\sigma^2)$ is calculated. If the number of minima $\#\mathcal{M}$ is equal to $C-1$, then the expected thresholds correspond to the detected minima and no further action is needed. Otherwise, the scale is increased (if $\#\mathcal{M}>C-1$) or decreased (if $\#\mathcal{M}<C-1$) in steps of $\Delta\sigma^2$ until the correct number of minima is obtained, and the resulting set of minima is returned as the threshold set $\mathcal{T}$; by the causality property recalled in Section~\ref{sec:Problem-formulation}, the number of minima is a non-increasing function of the scale, so the search direction is well defined. This procedure is summarized in
Algorithm~\ref{alg:fullprocedure}. The loop as printed assumes that the minima count passes through $C-1$; when $\Delta\sigma^2<0$, the search is in addition bounded by the positivity constraint $\sigma^2+m\Delta\sigma^2>0$. If the count jumps over $C-1$ within a single step, which can happen when two minima appear at nearly the same scale, the step $\Delta\sigma^2$ can be reduced and the search restarted from the previous scale.

\begin{algorithm}[!t]
\caption{Histogram clustering}
\label{alg:fullprocedure}
\begin{algorithmic}
\REQUIRE $\mathcal{H},C$
\STATE Find KD$(t,\mathcal{B},\sigma^2)$ using Algorithm~\ref{eq:EM-iterations}.
\STATE Find the set $\mathcal{M}$ of local minima of KD$(t,\mathcal{B},\sigma^2)$.
\IF{$\#\mathcal{M}=C-1$}
  \STATE Set $\mathcal{T}=\mathcal{M}$.
  \RETURN
\ELSIF{$\#\mathcal{M}>C-1$}
  \STATE Set $\Delta\sigma^2>0$.
\ELSE
  \STATE Set $\Delta\sigma^2<0$.
\ENDIF
\STATE Initialize $m=1$.
\REPEAT
  \STATE Update $\mathcal{M}$ as the set of local minima of KD$(t,\mathcal{B},\sigma^2+m\Delta\sigma^2)$.
  \STATE Set $m=m+1$.
\UNTIL{$\#\mathcal{M}=C-1$}
\STATE Set $\mathcal{T}=\mathcal{M}$.
\ENSURE thresholds $\mathcal{T}$.
\end{algorithmic}
\end{algorithm}

\begin{figure}[!t]
\centering{}\includegraphics[width=4in]{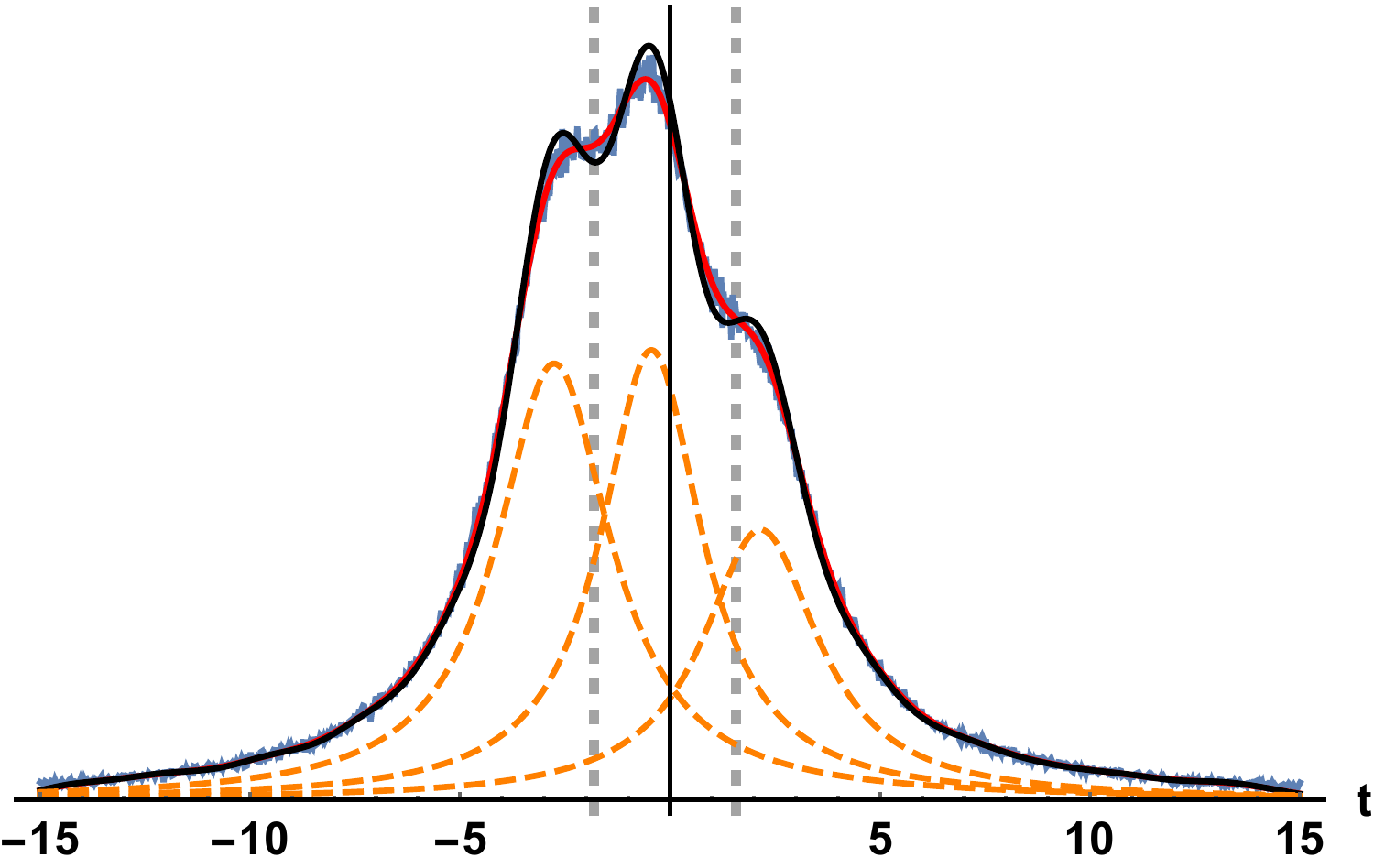}
\caption{KD estimation of a noisy histogram whose underlying distribution is a mixture of three modes. The blue solid line shows the histogram, the red solid line the KD
estimate, the dashed orange lines the modes of the underlying distribution, the solid black line the KD estimate at the scale at which the local minima are resolved,
and
finally the vertical dashed gray lines the threshold values extracted from the minima of the KD estimate.}
\label{fig:KD-estimation}
\end{figure}

Figure~\ref{fig:KD-estimation} shows the thresholds identified by the algorithm on a simulated histogram composed of three Gaussians plus additive noise. The detected thresholds almost coincide with the positions of the intersections of the different weighted Gaussians. An advantage of the proposed algorithm is its reduced sensitivity to the histogram bin width. A direct detection of the local minima is affected by the width $\Delta t$, since these minima may or may not be present at a given resolution. Since the proposed algorithm is based on a continuous scale-space representation of the original histogram, the local minima can instead be recovered at sub-bin resolution, largely independently of the discretization; Section~\ref{sec:Thresholding-synthetic} quantifies this for histograms with 100 and 1000 bins (Figure~\ref{fig:Comparison-with-truth}).

In Sections~\ref{sec:Thresholding-synthetic} and~\ref{sec:Thresholding-XCT}, we validate the proposed algorithm on synthetic histograms and on histograms obtained from XCT images, respectively.

\section{Validation on synthetic histograms}\label{sec:Thresholding-synthetic}
We validate the method against a set of 2185 synthetic histograms. The histograms are generated using a mixture of three Cauchy distributions defined by
\begin{equation}
F\left(t\right)=\sum_{k=1}^{3}\frac{c_{k}}{\pi b_{k}\left(1+\left(\frac{t-a_{k}}{b_{k}}\right)^{2}\right)},\label{eq:synthetic-histogram}
\end{equation}
where $c_{k}$ is the \textit{a priori} probability of the $k$-th component (the weights are normalized so that $\sum_k c_k=1$), and $a_{k}$ and $b_{k}$ are the location and width parameters of the Cauchy distribution. We generate a set of 2185 different distributions by uniformly selecting the mixture parameters with $-4<a_{k}<4,\ 0.5<b_{k}<2$ and $0.2<c_{k}<0.5$; the components are ordered by their location parameters $a_k$. Note that Cauchy components have no finite mean or variance, so this experiment violates the Gaussian mixture assumption of Section~\ref{sec:Problem-formulation} and thereby probes the robustness of the method with respect to model mismatch.

The reference thresholds for each parameter set $\{(a_{k},b_{k},c_{k})\}_{k=1}^{3}$ are the two intersection points of consecutive weighted mixture components, taken between their locations. Using each mixture distribution, we generate $10^4$ samples; these are used to create histograms with either 100 or 1000 bins in the interval $\left[-15,15\right]$. Next, we apply the proposed method to each synthetic histogram to obtain the corresponding thresholds and we compare them with the reference values. In all experiments, the histograms are processed on the bin-index grid (i.e., $\Delta t=1$ in Algorithm~\ref{eq:EM-iterations}) and we use the scale step $\Delta\sigma^2=0.01$ in these units; the resulting thresholds are mapped back to the intensity axis.

\begin{figure}[!t]
\begin{centering}
\subfloat[100 bins]{\begin{centering}
\includegraphics[width=3.5in]{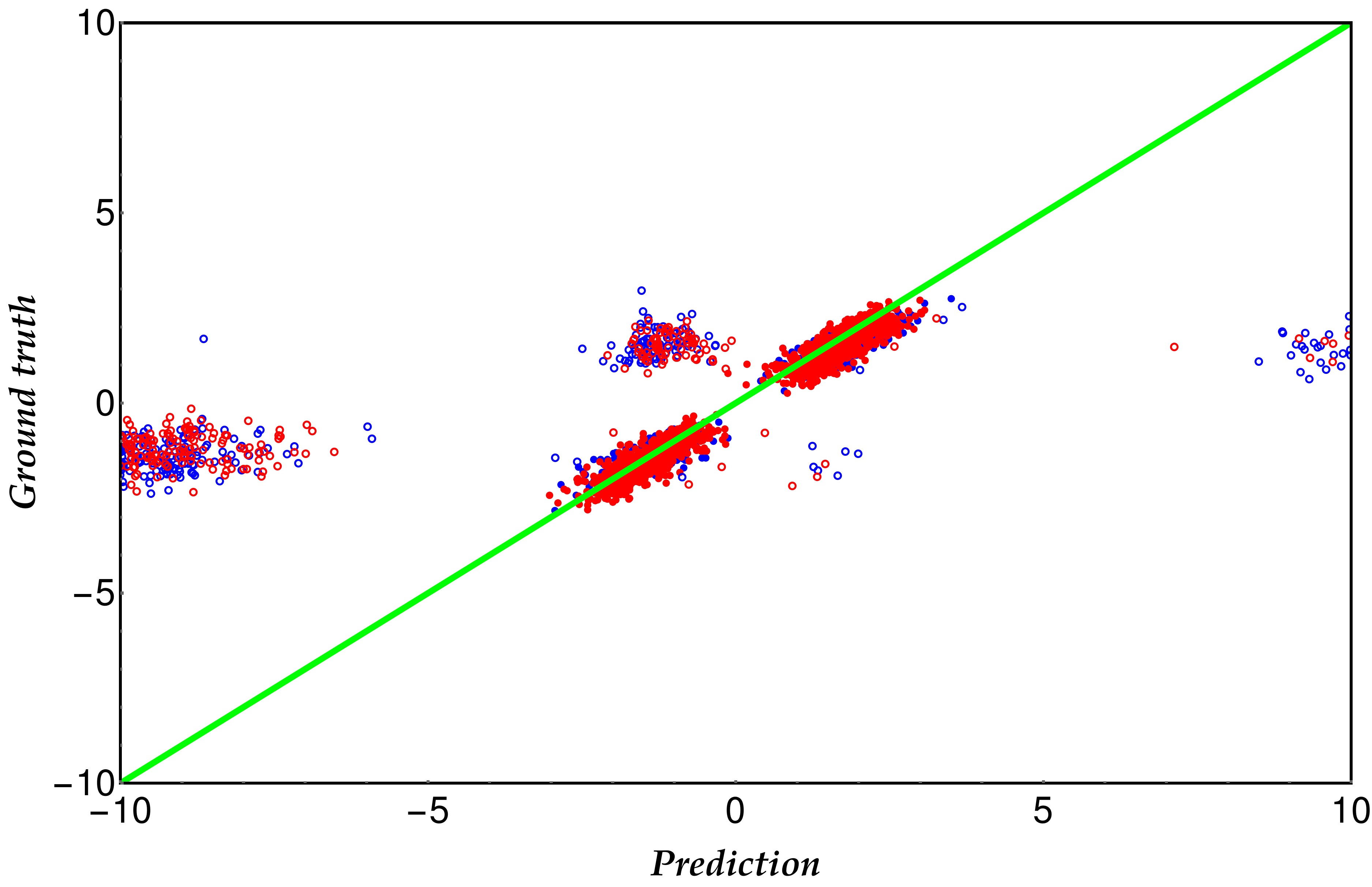}
\par\end{centering}
}\hfill{}\subfloat[1000 bins]{\begin{centering}
\includegraphics[width=3.5in]{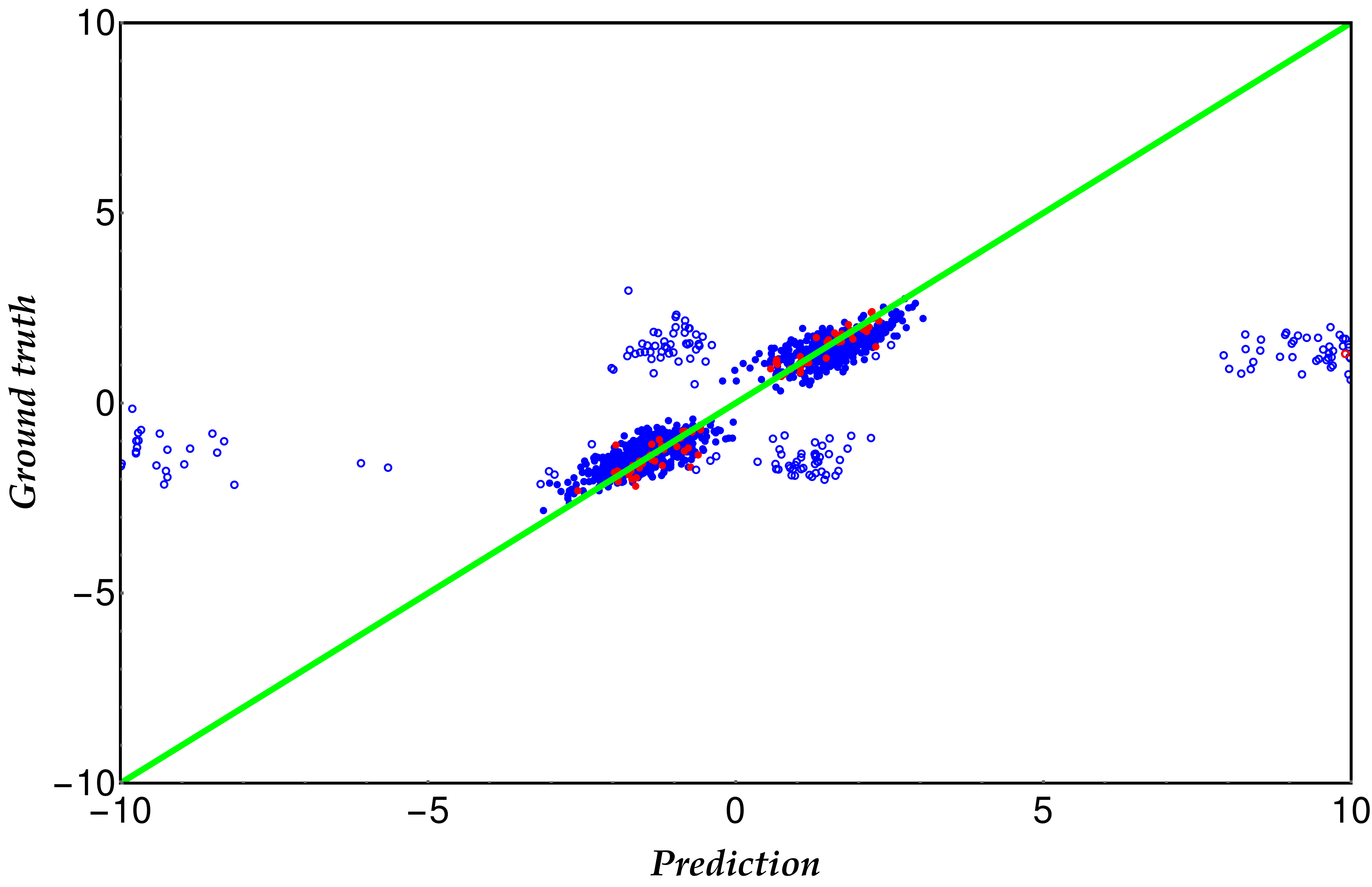}
\par\end{centering}
}
\par\end{centering}
\centering{}\caption{Comparison between the predicted ($x$ axis) and reference ($y$ axis) threshold values for synthetic histograms with (a) 100 bins and (b) 1000 bins. The blue and red markers refer to cases where the number of local minima of the KD estimate at the fitted scale is greater (i.e., $\#\mathcal{M}>C-1$) or smaller (i.e., $\#\mathcal{M}<C-1$) than the number of requested threshold values, respectively. The green line shows the function $y=x$. The filled markers correspond to threshold values that lie within the band $\left|x-y\right|<1$.}
\label{fig:Comparison-with-truth}
\end{figure}

Figure~\ref{fig:Comparison-with-truth} illustrates the accuracy of the predicted thresholds versus their reference values, where closeness to the $y=x$ line (solid green line in Figure~\ref{fig:Comparison-with-truth}) indicates a more accurate prediction. The filled markers correspond to threshold values that lie within the band $\left|x-y\right|<1$; the blue and red markers refer to cases where the number of local minima of the KD estimate at the fitted scale, before the scale adjustment, is greater (i.e., $\#\mathcal{M}>C-1$) or smaller (i.e., $\#\mathcal{M}<C-1$) than the number of requested threshold values, respectively. We find that about $22\%$ of the predictions based on 100-bin histograms deviate from the reference values by more than $1$ on the interval $\left[-15,15\right]$; for the histograms based on 1000 bins, this percentage decreases to $14\%$. Many of the largest deviations are predicted thresholds that lie far in the tails of the histogram ($|x|\approx 10$), consistent with the spurious tail minima discussed in Section~\ref{sec:Conclusion}. These experiments show that the proposed algorithm recovers the majority of the thresholds accurately, and that the accuracy improves with the histogram resolution.

\section{Application to porosity estimation from XCT images}\label{sec:Thresholding-XCT}
\begin{figure}
\subfloat[Full image\label{fig:phases-full}]{\includegraphics[width=0.49\textwidth]{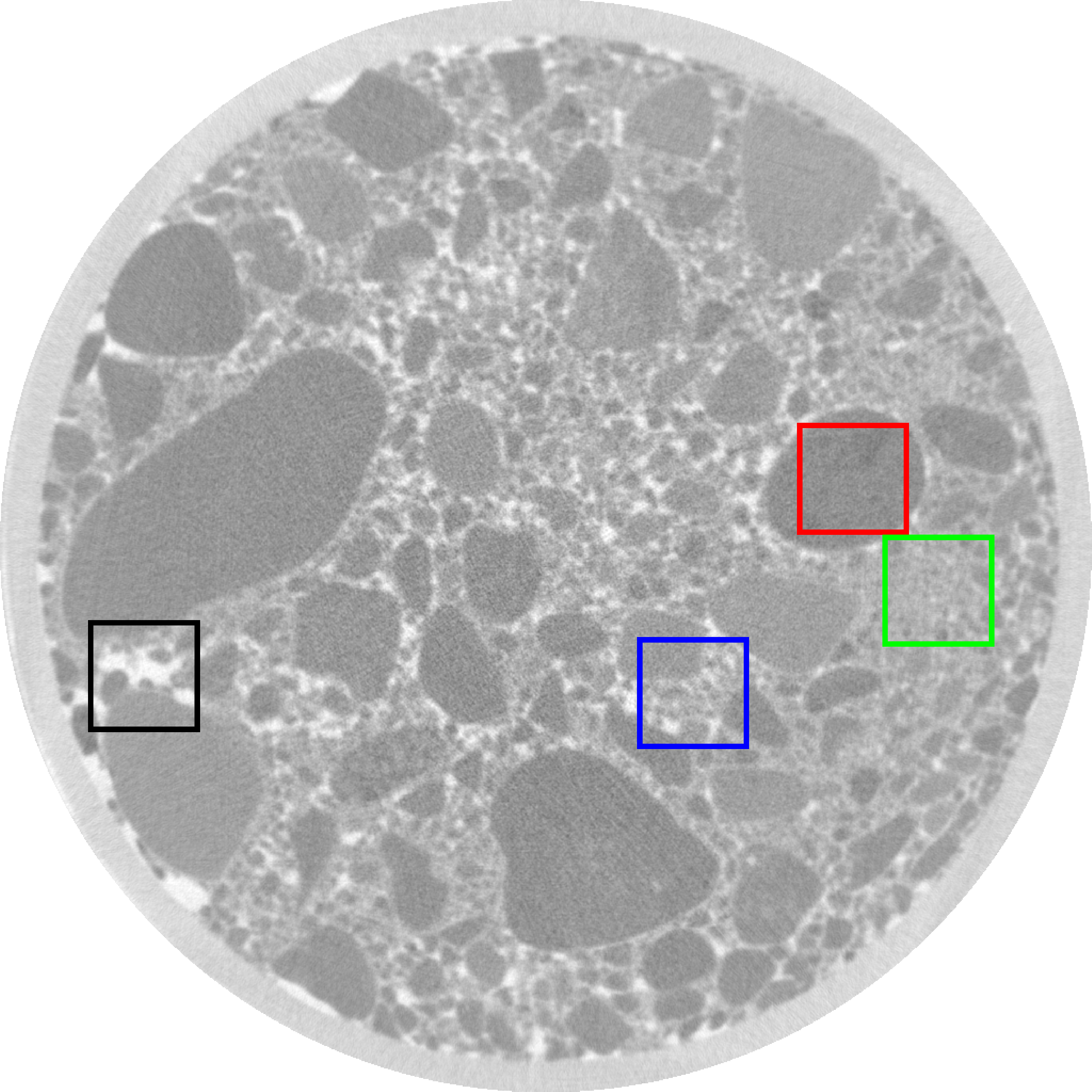}
}\hfill{}\subfloat[Zoomed-in view\label{fig:phases-zoom}]{\includegraphics[width=0.49\textwidth]{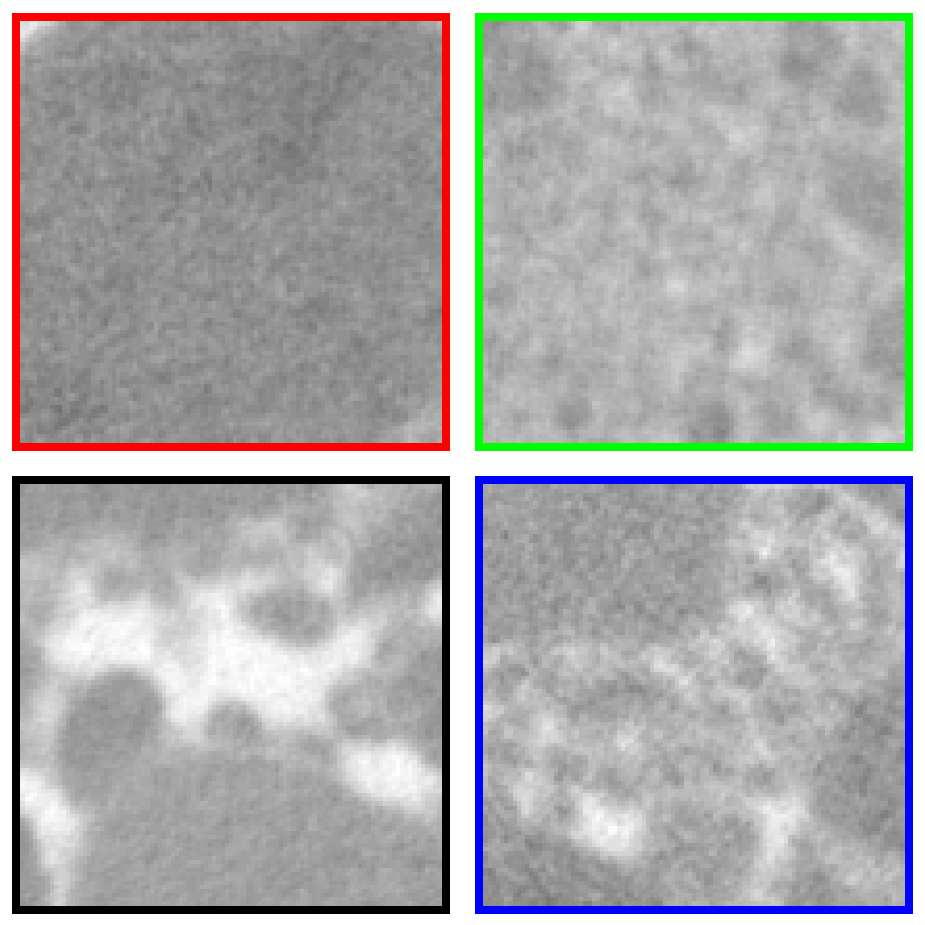}
}
\caption{(a) A representative XCT image of sediment column C6197A. (b) Zoomed-in
views of the solid phase (red square), the unresolved porous phase (green square),
the void phase (black square) and a mixture of the three phases (blue
square).\label{fig:phases}}
\end{figure}

In this section, we present an application of our algorithm to the estimation of the porosity of a natural porous sample from XCT images.
We apply the proposed thresholding algorithm to a stack of 1000 XCT gray-scale images of the sample; the algorithm is applied to a single histogram accumulated over
all pixels of all images in the stack. An example of such an image is given in Figure~\ref{fig:phases-full}. All details about the sample as well as the XCT imaging technique are provided in~\cite{Korneev2018}.

The pixel intensity of an XCT image is linearly related to the pixel porosity through the map~\cite{Korneev2018}
\begin{equation}\label{eq:porositymap}
\phi\left(t\right)=\dfrac{t-t_{s}}{t_{v}-t_{s}},
\end{equation}
where $t$ is the pixel intensity, and $t_s$ and $t_v$ are two reference points that have to be determined. The reference point $t_s$ corresponds to the pixel intensity of the solid phase ($\phi=0$), and $t_{v}$ to the pixel intensity of the void phase ($\phi=1$). Such reference points are identified by thresholding the histogram. From the properties of the XCT imaging and of the sample, the combined image intensity histogram is known to consist of three clusters: the first cluster (lowest intensities) represents the void phase, the second cluster the unresolved porous phase, and the third cluster the solid phase~\cite{Korneev2018}. The images in Figure~\ref{fig:phases} are displayed with an inverted gray scale, so the void phase appears white and the solid phase dark gray (see the zoomed-in views in Figure~\ref{fig:phases-zoom}).

\begin{figure}
\subfloat[Histogram\label{fig:histogram-real-original}]{\includegraphics[width=0.49\textwidth]{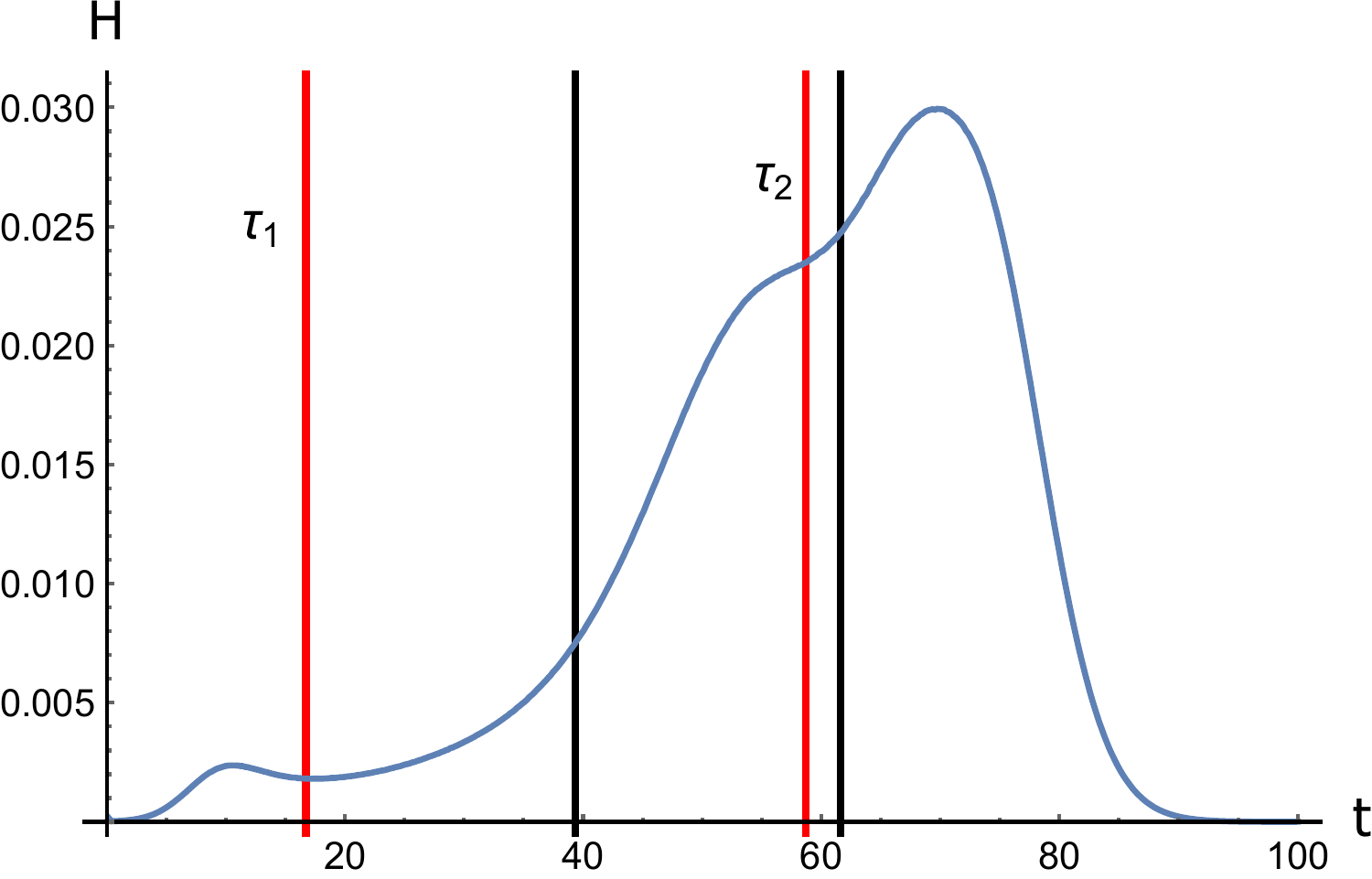}
}\hfill{}
\subfloat[Deconvolved histogram\label{fig:histogram-real-deconv}]{\includegraphics[width=0.49\textwidth]{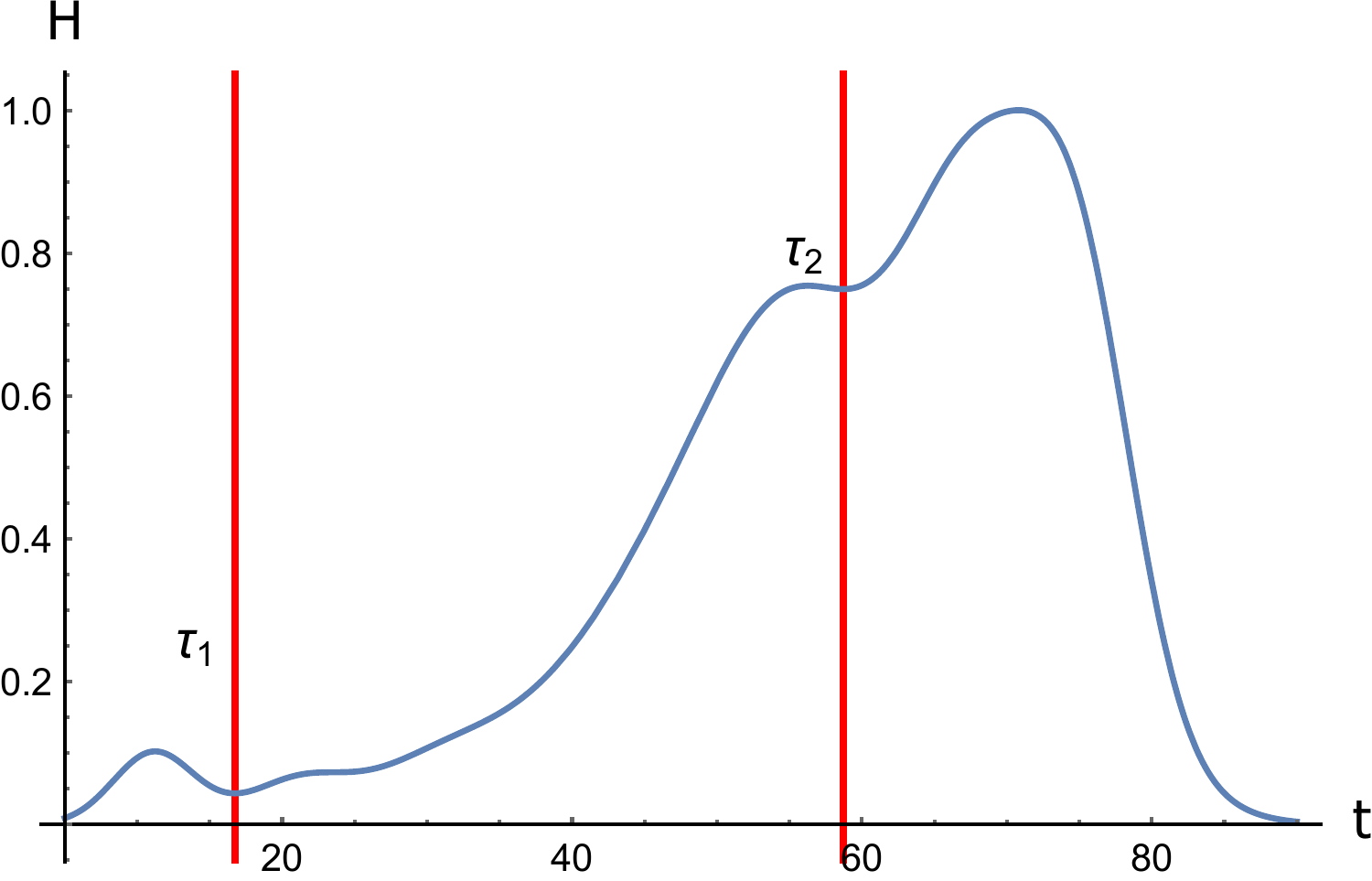}
}
\caption{The combined intensity histogram (blue line) (a) at the original scale, and (b) at the scale at which the minima are resolved. The vertical red and black lines indicate the threshold values extracted by the proposed method and by K-means clustering, respectively. \label{fig:histogram-real}}
\end{figure}

The combined histogram, shown in Figure~\ref{fig:histogram-real-original}, has three highly overlapping modes: at the original scale, only the minimum separating the void and porous phases is visible, while the minimum separating the porous and solid phases is not resolved. We apply the proposed method (with $\Delta\sigma^2=0.01$) to the histogram and obtain the thresholds $\tau_1$ and $\tau_2$ as indicated in Figure~\ref{fig:histogram-real-deconv}. Given $\tau_1$ and $\tau_2$, the reference points $t_v$ and $t_s$ are determined as the mean intensities within the first and third clusters, respectively, i.e.,
\begin{align}
t_{v}=\frac{\sum_{t_{i}<\tau_{1}} h_{i}t_{i}}{\sum_{t_{i}<\tau_{1}} h_{i}},\qquad
t_{s}=\frac{\sum_{t_{i}>\tau_{2}} h_{i}t_{i}}{\sum_{t_{i}>\tau_{2}} h_{i}}.
\end{align}
Next, the gray intensity histogram is transformed into a porosity histogram via \eqref{eq:porositymap}. The average porosity is then calculated by computing the first moment of the porosity histogram. The estimated and experimentally measured porosities are $0.185$ and $0.178$, respectively, with a relative error of $3.9\%$. For comparison, we follow the same procedure using the thresholds obtained by the classic K-means clustering algorithm. The obtained porosity value in this case is $0.27$, which corresponds to a relative error of $51.6\%$.

\section{Conclusions}\label{sec:Conclusion}

In this paper, we introduced a new approach to segment histograms whose modes overlap and are not sufficiently resolved. The algorithm exploits the properties of scale-space representations to automatically find the appropriate thresholds. Its main parameters are the number of expected classes $C$ and the scale step $\Delta\sigma^2$; in addition, the EM estimation of the kernel bandwidth uses a stopping tolerance $\delta$, which acts as an implicit regularization parameter. We first tested the proposed method on a large set of synthetic histograms. We then applied the algorithm to the estimation of the porosity of a natural porous sample from XCT images and compared its performance with the K-means algorithm.

The proposed algorithm has an intuitive formulation for 1D histograms, but its extension to multivariate histograms is not straightforward. The generalization of the deconvolution part is simple, but the definition of the thresholding surfaces is a challenging problem. The local minima criterion does not carry over to a multivariate histogram: a multidimensional function may have a very complex structure of extrema, including saddle points and long, narrow valleys~\cite{Carreira2003}. Another weakness of the algorithm is associated with the use of a mixture of Gaussians with a common variance. This type of estimate represents histograms with long flat tails poorly and may produce spurious local minima in the flat part of the histogram, which is consistent with the largest deviations observed in Section~\ref{sec:Thresholding-synthetic}.

The strength of the algorithm is its ability to separate highly overlapping clusters, including clusters that have merged into a unimodal distribution. Classical multilevel criteria, such as Otsu's method~\cite{Otsu1979,Liao2001} or minimum error thresholding~\cite{Kittler1986}, also return $C-1$ thresholds on a unimodal histogram, but they rely on a global variance-based or parametric criterion, whereas the proposed method requires no per-class model and recovers the merged valleys directly; clustering algorithms generally require the clusters to be well separated. Existing methods designed specifically for unimodal histograms~\cite{Rosin2001,Coudray2010,Medina-Carnicer2008} threshold them into two clusters only, while the proposed method can threshold a unimodal histogram into more than two classes. The deconvolution also reduces the sensitivity of the extracted minima to the choice of the bin width: local minima may appear or disappear depending on the bin width, whereas the continuous KD model extracts the hidden local minima, and convolution smooths out the minima which are less significant. The proposed algorithm is computationally efficient compared to the direct clustering of the data: clustering a data set consisting of billions of points is demanding for a classical algorithm, while the calculation of a histogram requires little computational effort and is easily parallelizable. Finally, the early stopping of the EM deconvolution makes the algorithm robust to high-frequency noise in the histogram shape.

Future work includes the extension of the algorithm to multivariate histograms, which requires the formulation of a new set of criteria defining the thresholding surfaces. Current investigations are focused on approaches to automatically identify the ``optimal'' number of classes, and on reducing the number of algorithm parameters.

\section*{Contributions}
The method, the algorithms, all experiments (the synthetic study of Section~\ref{sec:Thresholding-synthetic} and the XCT porosity application of Section~\ref{sec:Thresholding-XCT}), and the original manuscript (arXiv v1, 2022) are the work of the listed authors, S.~Korneev, J.~Gilles and I.~Battiato.

The present revision (v2) was prepared with substantial assistance from Claude Fable~5, an AI assistant developed by Anthropic (San Francisco, CA, USA). The assistant: identified and corrected two defects in the printed algorithms of v1 (the inverted termination condition of Algorithm~\ref{alg:fullprocedure}, and the bandwidth update of Algorithm~\ref{eq:EM-iterations}, which carried a spurious $\sigma^{2}$ factor and lacked the discrete-convolution conventions needed to reproduce it); restated the estimation objective as the Csisz\'ar I-divergence that the EM iterations minimize, together with the ill-posedness of the joint estimation and the role of early stopping; revised the statements relating mixture minima to minimum-Bayes-error thresholds and the scale-space causality discussion; added the positioning with respect to critical-bandwidth selection, mode trees, multilevel thresholding criteria and related work \cite{Silverman1981,Minnotte1993,Otsu1979,Liao2001,Barron2020,Babaud1986,Carreira2003,Lucy1974}; verified the corrected algorithm numerically; and performed a full editorial revision. The assistant contributed no new experiments; all data, results and scientific claims are those of the original study. Following arXiv's policy on generative AI tools, the assistant's role is disclosed here rather than through authorship.

\section*{Acknowledgments}
I. Battiato and S. Korneev have been supported by the U.S. Department of Energy Early Career Award in Basic Energy Sciences DE-SC0019075. The v2 revision was prepared with the assistance of Claude Fable~5 (Anthropic); see the Contributions section.

\bibliographystyle{abbrv}
\bibliography{paper}

\end{document}